 \documentclass[sigconf]{acmart}
\usepackage{soul}
\usepackage{longtable}
\usepackage[resetlabels,labeled]{multibib}
\newcites{R}{Sampled Work}

\AtBeginDocument{%
  }

\copyrightyear{2023}
\acmYear{2023}
\setcopyright{rightsretained}
\acmConference[CHI '23]{Proceedings of the 2023 CHI Conference on Human Factors in Computing Systems}{April 23--28, 2023}{Hamburg, Germany}
\acmBooktitle{Proceedings of the 2023 CHI Conference on Human Factors in Computing Systems (CHI '23), April 23--28, 2023, Hamburg, Germany}\acmDOI{10.1145/3544548.3580962}
\acmISBN{978-1-4503-9421-5/23/04}

\begin{document}

\title{A Framework and Call to Action for the Future Development of EMG-Based Input in HCI}

\author{Ethan Eddy}
\email{eeddy@unb.ca}
\orcid{0000-0002-8392-3729}
\affiliation{%
  \institution{University of New Brunswick}
  \city{Fredericton}
  \country{Canada}
}

\author{Erik Scheme}
\email{escheme@unb.ca}
\orcid{0000-0002-4421-1016}
\affiliation{%
  \institution{University of New Brunswick}
  \city{Fredericton}
  \country{Canada}
}

\author{Scott Bateman}
\email{scottb@unb.ca}
\orcid{0000-0003-3592-2163}
\affiliation{%
  \institution{University of New Brunswick}
  \city{Fredericton}
  \country{Canada}
}


\renewcommand{\shortauthors}{Eddy, Scheme, and Bateman}

\begin{abstract}
Electromyography (EMG) has been explored as an HCI input modality following a long history of success for prosthesis control. While EMG has the potential to address a range of hands-free interaction needs, it has yet to be widely accepted outside of prosthetics due to a perceived lack of robustness and intuitiveness. To understand how EMG input systems can be better designed, we sampled the ACM digital library to identify limitations in the approaches taken. Leveraging these works in combination with our research group's extensive interdisciplinary experience in this field, four themes emerged (1) interaction design, (2) model design, (3) system evaluation, and (4) reproducibility. Using these themes, we provide a step-by-step framework for designing EMG-based input systems to strengthen the foundation on which EMG-based interactions are built. Additionally, we provide a call-to-action for researchers to unlock the hidden potential of EMG as a widely applicable and highly usable input modality.

\end{abstract}



\maketitle

\section{Introduction}
The HCI community is constantly searching for new hands-free input modalities to enable more intuitive, convenient, and efficient input for interactive systems. One such method is through the use of electromyography (EMG) signals – the electrical impulses produced during muscular contractions \cite{yousefi_characterizing_2014}. EMG is a particularly attractive solution as it provides a compact and low-powered means for detecting muscle-based inputs while avoiding many of the limitations imposed by other technologies. For example, it eliminates the need for cumbersome equipment such as data gloves \cite{bhaskaran_smart_2016} and can be less invasive and computationally complex than computer vision approaches \cite{cheng_survey_2016, wang_superpixel-based_2015}. Additionally, depending on the desired interaction, users may elicit low-level muscle contractions that are faster and more subtle than the dynamic gestures or body postures required for other input modalities. While different sensing methods exist for measuring EMG signals \cite{bakiya_information_2018, merletti_surface_2001}, \textbf{surface EMG} is by far the most common and attractive method used today as it is non-invasive and relatively robust.

Surface EMG (sEMG) uses electrodes placed directly on the surface of the skin to measure the electrical activity over a muscle site. The EMG signals acquired from these sensors are then converted into control signals that can be used to control a device or interface, effectively leveraging natural human contractions as an intuitive control input. sEMG technology has been around since the early 1900s \cite{oldest_prosthetic1} but has remained largely inaccessible for general-purpose use due to a lack of inexpensive, usable, and commercially available equipment. 

The evolution of myoelectric control --- the control of a device using EMG signals \cite{asghari_oskoei_myoelectric_2007} --- has primarily been driven by its niche application in prosthetics, enabling amputees to operate a powered prosthesis using the muscles of their residual limb. Early myoelectric control systems took a \textbf{one-muscle one-function approach}, where antagonistic muscles --- muscles whose actions oppose each other (e.g., flexion/extension of the wrist) --- were used as inputs for device control \cite{scheme_electromyogram_2011}. Each muscle was the functional equivalent of a momentary switch, with outputs being turned on when the muscle contraction exceeded some threshold. While this resulted in a simple control scheme, it offered a limited number of viable sites for control inputs and thus required unintuitive mode switching to increase the input space and enable effective device control \cite{farina_extraction_2014}. To improve the control of more degrees of freedom, the synergistic behaviour between multiple muscle sites was introduced using \textbf{pattern recognition} \cite{englehart_robust_2003}. Pattern recognition leverages machine learning algorithms to learn and detect repeatable patterns in EMG signals. Over the past two decades, the prosthetics community has slowly improved pattern recognition-based myoelectric control, reaching near-perfect offline classification accuracies (in controlled lab-based settings) \cite{benalcazar_real-time_2018, zhang_framework_2011,leon_emg_2011}, and pattern recognition-based myoelectric prosthesis control is now commercially available from several vendors \cite{coapt_web, ottobock_web, biomed}. Nevertheless, the research community continues to explore improvements to its intuitiveness, robustness, and dexterity, such as through algorithms that enable simultaneous independent proportional control (control over multiple degrees of freedom at once) \cite{hahne_linear_2014, ameri_support_2014, jiang__extracting_2009}. Despite this rich history of success in prosthesis control, however, it wasn't until the mid-2000s that EMG began to gain popularity within the HCI community \cite{costanza_intimate_2007,costanza_toward_2005,saponas_demonstrating_2008,saponas_making_2010, saponas_enabling_2009,naik_hand_2006}. 

With the release of an inexpensive commercially available sEMG device in 2014 \cite{rawat_evaluating_2016}, EMG became a viable option for broad use by HCI researchers. Since then, EMG has been leveraged for a wide range of general-purpose applications, such as: controlling drones \cite{stoica_remote_2014}, enhanced piano control \cite{karolus_hit_2020}, sign language recognition \cite{paudyal_sceptre_2016}, RC car control \cite{kim_emg-based_2008}, spelling systems \cite{vasiljevas_development_2014}, interactions with games \cite{zhang_hand_2009}, and augmenting gym experiences \cite{ho_myobuddy_2017}, to name a few. However, while EMG-based control has become accessible to the HCI community and is continuing to grow in popularity, it has yet to make significant strides toward being used in real-world general-purpose applications. By `general-purpose' we mean applications that are outside the most common use of controlling a prosthetic limb. 

\subsection{Scope and Contribution}
To increase success in research and development practices for using EMG-based inputs in general-purpose applications, we make three specific contributions. First, we synthesize the main limitations of EMG-based input that exist in a sampling of HCI work. Second, we outline a framework for designing and building EMG-based input control in interactive systems. Third, we provide a call to action identifying the key areas of research that need to be addressed for the continued and improved success of EMG-based input. We elaborate on these contributions below.

The lack of widespread adoption of EMG-based control may be attributed, at least in part, to the fact that designing and implementing robust EMG-based input systems is inherently complex and requires careful treatment of the stochastic EMG signal. As a result, the successful adoption and application of EMG-based control for HCI researchers as part of a larger interactive system is extremely challenging. Inevitably, this has led to systems lacking the performance, robustness, and intuitiveness required for everyday, real-world use in general-purpose applications. These observations corroborate those made by members of our larger research team and through our longstanding collaborations with partners in both the prosthetics community and HCI fields. Our research team includes members of an HCI research lab, a biomedical engineering research lab, and a prosthetics clinic. We have published EMG research in HCI and biomedical engineering venues, in addition to working with industry collaborators who are developing commodity-level general-purpose EMG wearables. These experiences have reinforced the notion that understanding the exact source and nature of the difficulties that are currently hindering the adoption of EMG-based control is crucial for its successful uptake as an enabling technology, beyond its use in powered prostheses. 

To characterize and synthesize the current challenges and limitations hindering the reliable use of EMG as an input modality outside of prosthesis control, we sampled the ACM library for HCI applications leveraging EMG for control. Through this process, we found an overarching theme of direct, and sometimes naive, adoption of the work done in the prosthetics community. Because this work has evolved for a highly specific --- and different --- use case, the reliance on methods designed specifically for prosthesis control may be leading to sub-optimal solutions within HCI and in applications to different contexts. Our findings suggest that for EMG systems to reliably perform at the levels that would be expected for real-world commercial applications, it is imperative that a solid basis and understanding of EMG and its use is available. From capturing EMG signals to the use of static contraction or dynamic gesture recognition, to algorithm selection, our findings suggest that HCI researchers could benefit from a common starting point that bridges the gap from prosthesis control to general-purpose use. While HCI researchers can leverage the expertise developed within prosthetics research, we must consider the unique challenges, opportunities, and use cases for EMG-based interactions with general-purpose applications in mind. 

Having identified potential limitations to the general-purpose use of EMG input and drawing on our experience in building EMG-based input from the ground up, we propose a design framework for developing EMG-based control systems, specifically with HCI researchers in mind. Further, we provide a roadmap and call to action for research into EMG-based input. In doing so, we hope to facilitate the design process and success of these systems, and in turn, challenge researchers to unlock the hidden potential of EMG for general-purpose applications.

\subsection{Layout}
The rest of the paper is divided into five main sections. Section \ref{sec:section2} (Background) provides a brief overview of the physiology of EMG and the history of its use for prosthesis control and general-purpose applications. This section sets the stage for exploring the use of EMG later in the paper. Section \ref{sec:section3} (Exploring EMG Applications in HCI) highlights the process that we followed to extract the current challenges and limitations hindering EMG-based development from an HCI perspective. We also discuss the four main themes from our findings: (1) interaction design, (2) model design, (3) system evaluation, and (4) reproducibility. Sections \ref{sec:section4} and \ref{sec:section5} (A Framework for Building EMG-Based Interactions and Further Challenges and Considerations) lay out a new design framework for designing and adopting these interactions. Finally, in section \ref{sec:section6}, we discuss this work and its limitations, and evaluate the future of EMG research, providing a call to action for the HCI community. 

\section{Background} \label{sec:section2}

\subsection{The EMG Signal}
The electromyography (EMG) signal is a manifestation of the electrical activity that is created by a muscle when it contracts \cite{phinyomark_surface_2020}. The initial discovery of these biological signals can be traced back to the mid-1600s \cite{criswell_crams_2010}, however, it was not until the early 1900s that advancements in technologies such as the galvanometer and the cathode ray oscilloscope enabled a more detailed analysis of the EMG signal \cite{criswell_crams_2010}. Since then, EMG has been used for many different purposes, including identifying neuromuscular diseases \cite{elamvazuthi_electromyography_2015,stalberg_scanning_1991,subasi_classification_2013}, general human-machine interaction \cite{assad_biosleeve_2013, karolus_hit_2020, paudyal_sceptre_2016, karolus_emguitar_2018}, and powered prosthesis control \cite{scheme_electromyogram_2011, parker_myoelectric_2006}. Physiologically, when a muscle contracts, action potentials propagate along the membrane of each muscle fiber. The combination of these action potentials from the different muscle fibers of each motor unit is known as a motor unit action potential (MUAP) \cite{raez_techniques_2006}. The EMG signal is the summation of all of these MUAPs and is dependent on the physiological properties of the particular muscle and the relative recording location \cite{raez_techniques_2006}. These action potentials occur at pseudo-random intervals and at varying distances from the sensors, causing the stochastic, or random, behaviour of the EMG signal \cite{phinyomark_surface_2020}. For general-purpose applications, sEMG is predominately used (as opposed to intramuscular EMG, which involves the insertion of wires, needles, or sensors into the muscle transcutaneously) because it is non-invasive, easy to use, and relatively robust \cite{scheme_electromyogram_2011}. Although preferable, sEMG remains susceptible to signal corruption due to motion-artifact, improper skin-electrode contact, and other transient factors \cite{young_effects_2011, scheme_electromyogram_2011, campbell_current_2020}. Nevertheless, the potential of sEMG as a solution for hands-free control remains a compelling option for computer interaction.

\subsection{EMG for Prosthesis Control}
With increasing technological capabilities for sEMG came the exploration of its potential use for prosthesis control in the early 1940s \cite{englehart_robust_2003}. As a direct result, the evolution of control schemes was guided by the clinical needs of amputees and their desire to impart continuous control over their prostheses. The initial systems (i.e., conventional/direct control) took a one-muscle one-function approach for device control \cite{scheme_electromyogram_2011}. In these control schemes, the EMG signal is typically measured from two main muscle sites (e.g., for transradial amputees, over the forearm flexor and extensor muscles) and their amplitudes are compared to predefined thresholds to determine whether to activate a prosthesis function in one direction or the other (e.g., close/open hand). The speed of the device may be controlled based on the intensity of contraction (\textbf{proportional control}), or a fixed speed may be adopted (\textbf{constant control}) \cite{fougner_control_2012}. When using conventional control, mode switching --- activated by the simultaneous contraction of antagonist muscle sites, such as the aforementioned flexor and extensor sites --- is required to control more than a single degree of freedom, or function. In this way, mode switching is used to cycle through a list of prosthesis functions (e.g., hand open/close, wrist rotation, etc.) to select which is activated by the direct control. The inadvertent \textbf{false activation} of this mode switch often occurs, resulting in erroneous control inputs causing unintended device selection or action \cite{tabor_quantifying_2017}. The unintuitive and sequential nature of these switches is also cumbersome. For example, imagine trying to control a cursor while having to co-contract each time to change control between the horizontal and vertical axes. This problem is exacerbated by every additional degree of freedom to be controlled within a prosthesis. As a result, prosthetics researchers turned to machine learning techniques to extract additional information and enable the control of more functions \cite{farina_extraction_2004}. 

Initial pattern recognition-based myoelectric control systems were developed out of the necessity for more intuitive and robust device control in the mid to late 1900s \cite{englehart_robust_2003}. Unlike conventional control, pattern recognition approaches use multiple EMG sites, \textbf{feature extraction techniques} (see section \ref{sec:Parameters}), and robust \textbf{algorithms} (see section \ref{sec:AlgSelection}) to leverage the added information in the way muscles work together \cite{scheme_electromyogram_2011}. Doing so removes the theoretical ceiling on the number of potential control outputs, as they are no longer tied to the number of electrode sites but, instead, the number of distinct patterns of contractions \cite{hudgins_new_1993}. In 2003, Englehart et al. set the standard for continuous pattern recognition-based myoelectric control still used today in commercial systems \cite{englehart_robust_2003}. A continuous stream of class decisions is sent to the prosthetic device and converted to a prosthesis function. This decision stream consists of \textit{N} known classes to which the current pattern of EMG corresponds and when combined with proportional control, is used to control the velocity of the associated prosthesis function. As a result, the prosthesis is constantly reacting to EMG activity, and thus device control requires dedicated cognitive effort \cite{godfrey_synergy-driven_2013}. Nevertheless, with a lot of training and practice (sometimes years), amputees can become extremely efficient using this control scheme \cite{hargrove_myoelectric_2017}.

\begin{figure*}[h!]
  \centering
  \includegraphics[width=14cm]{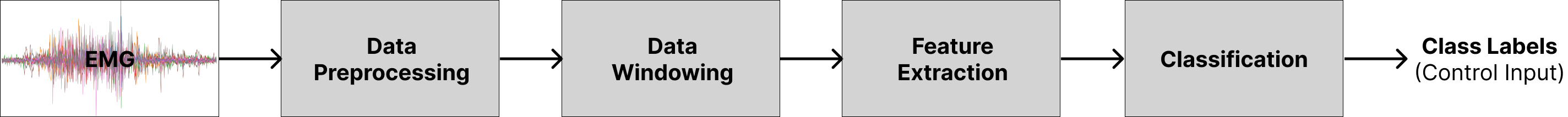}
  \caption{The stages of continuous EMG-based pattern recognition for prosthesis control. }
  \Description{The steps required for pattern recognition-based recognition, starting with EMG as the input and ending with class labels as outputs. Every step is displayed in a box with arrows progressing the timeline forward.}
  \label{fig:PR_EMG_Steps}
\end{figure*}

As exemplified in \cite{scheme_electromyogram_2011}, pattern recognition-based EMG control systems used today for prosthesis control consist of the main stages shown in Figure ~\ref{fig:PR_EMG_Steps}. In the data preprocessing stage, unwanted sources of noise are filtered from the raw EMG signal. Filtering noise is important as it can negatively impact the ability of algorithms to differentiate between patterns of contractions. Next, the filtered data are divided into fixed-length overlapping windows (or frames) from which descriptive features are extracted. These features are necessary to overcome the stochastic behaviour of EMG and to increase the information density of the signal prior to classification. Based on the extracted features, a machine learning model classifies each window of EMG data as belonging to one of the specified muscle-based inputs (e.g., hand open). The corresponding decisions are constantly output as a series of class labels, each corresponding to a particular muscle-based input, and ultimately converted into prosthesis control commands. These windows are kept very short (on the order of milliseconds) to enable quick and responsive control of the prosthesis. Finally, if using proportional control, the decision stream can be combined with the amplitude of the EMG-based signal to control the velocity of the device \cite{scheme_motion_2014}. As these classification-based pattern recognition approaches have become commercialized and clinically adopted, researchers have increasingly turned their attention to more complex control schemes.

A major limitation of classification-based approaches is that they require the sequential activation of motion classes (e.g., first wrist flexion, then wrist rotation, then hand open). Although combined classes can be incorporated (e.g., a class for both wrist flexion and wrist pronation at the same time) \cite{young_classification_2013}, the combinations of classes rapidly increase the training requirements for the user and the classifier. Consequently, alternative approaches - such as regression \cite{hahne_linear_2014, ameri_support_2014} - have been explored, which continuously map the input EMG to output activations in multiple degrees of freedom, enabling the simultaneous and independent proportional control of multiple functions \cite{ameri_real-time_2014}. Although such approaches offer the potential of added dexterity for prosthesis control, they fall within the same control paradigm of continuous control, which may not always apply in real-world HCI applications. 

\subsection{EMG in HCI}
While EMG-based control has been adopted clinically for prosthesis control for decades, it only began to garner attention from the HCI community for general-purpose use in the early 2000s \cite{costanza_intimate_2007,costanza_toward_2005,saponas_demonstrating_2008,saponas_making_2010, saponas_enabling_2009,naik_hand_2006}. This timing coincides with many prosthesis-related EMG control achievements, including increased attention due to an influx of funding via the US Defense Advanced Research Projects Agency's (DARPA) Revolutionizing Prosthetics program \cite{miranda_darpa-funded_2015}. Then, in 2008, Saponas et al. articulated the need for a commercially available wearable device to enable practical and inexpensive use of EMG for everyday use in HCI \cite{saponas_demonstrating_2008}. It wasn't until 2014, though, that the Myo Armband became arguably the first notable commercially available sEMG device \cite{rawat_evaluating_2016}.

\begin{figure}[h]
  \centering
  \includegraphics[width=4cm]{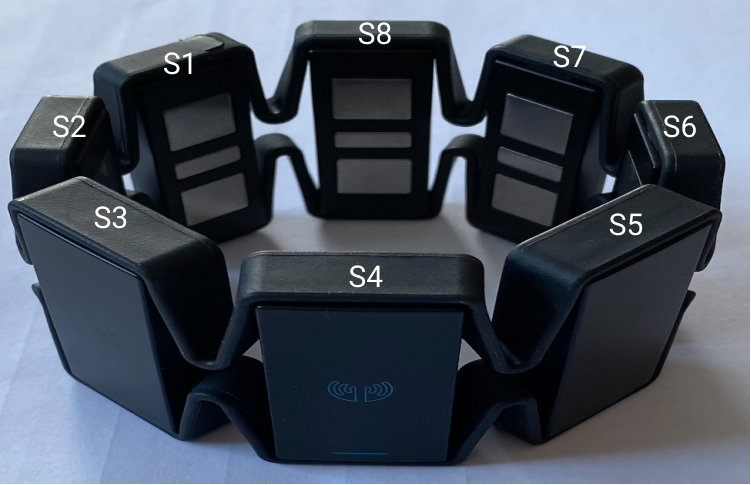}
  \caption{Myo Armband – Each S corresponds to an individual bipolar differential electrode sensor.}
  \Description{The Myo Armband is displayed with each sensor highlighted. In total there are eight different sensors that make up the armband.}
  \label{fig:MyoArmband}
\end{figure}

The Myo Armband (Figure ~\ref{fig:MyoArmband}) was a low-cost commercially available sEMG wearable band. The advantage of the Myo was that it was an inexpensive, convenient, and relatively robust method for obtaining sEMG data. The armband consisted of eight bipolar pairs of surface electrodes (8 channels) that measured EMG activity at 200 Hz \cite{koskimaki_myogym_2017}. While this was below the typical 1000-2000 Hz sampling rates of medical-grade electrodes, it was adequate for the reliable detection of many gestures and kept cost low \cite{phinyomark_feature_2018}. Additionally, the armband had a built-in 9-axis inertial measurement unit (IMU), potentially enabling the recognition of more dynamic inputs \cite{campbell_differences_2020, mccullough_imu}. The data were streamed wirelessly using Bluetooth Low Energy (BLE) to any BLE-enabled computing device. Although the Myo product has since been discontinued \cite{tortora_dual-myo_2019, prahm_developing_2022}, it set the stage for enabling EMG as a new input modality for general-purpose applications. 

With the rapid evolution of virtual and augmented reality, there is growing interest in novel interaction techniques that do not rely on physical devices, like mice or keyboards, as controllers. As such, there has been a renewed interest in the adoption of EMG technology in research (e.g.,\cite{hu_comprehensive_2020, dai_capg-myo_2021, zhang_intelligent_2022}) and commercial applications (e.g.,  \cite{jaloza_inside_2021}). However, we believe that for the successful adoption of EMG in HCI, we must deepen our understanding of its use and define new directions for improved systems that are not explored meaningfully in the body of prosthetics research. Leveraging EMG for general-purpose applications has its own unique challenges and intricacies that must be considered and explored to ensure its future success.

\section{Exploring EMG Applications in HCI}
\label{sec:section3}
In this section, we analyze previous EMG-based control work in the HCI literature to identify the main challenges and limitations. Despite the individual contributions of many works, four main themes emerged that may be limiting the advancement of the field: (1) \textbf{interaction design}; the improper or sub-optimal selection of a control scheme for the desired interaction, (2) \textbf{model design}; improper design or optimization of the components of the control scheme, (3) \textbf{system evaluation}; evaluation of system performance in isolation or without sufficient interaction context, and (4) \textbf{reproducibility}; insufficient experimental control or detail provided to enable the replication, corroboration, or extension of presented results. 

\subsection{Paper Selection Process}
Over the past two decades, the HCI community has contributed creative and valuable research in the area of EMG-based control. Even with the growing academic popularity of EMG, however, it has yet to make significant strides toward reliable commercial general-purpose use. Consequently, the goal of this sampling was to highlight the main barriers and challenges to EMG-based control for use in HCI research and consumer applications. We correspondingly sampled papers exclusively from the Association for Computing Machinery (ACM) library to focus on the HCI community and shift the focus away from the comparatively much larger and older EMG-based prosthesis control literature (predominantly found in IEEE or clinical journals). From the ACM body of literature, we selected approximately 50 papers on the use of EMG for general-purpose applications. We provide references for these specific papers in an additional bibliography (see \hyperref[ReviewedWork]{Sampled Work}). The process used for paper selection was rather lenient but focused on those related to using EMG as an input for the control of an interactive system. This was therefore not meant to be a formal systematic review, but instead, a sampling to characterize the main challenges and limitations that HCI researchers are currently facing, drawing on a combination of the available literature and our collective experiences in both prosthetics and HCI. From each of the selected papers, we documented the following categories of information: application/context, inputs, algorithms(s), features used, windowing technique, training methods, user/context dependence, evaluation technique(s), evaluation metric(s), reported accuracy, number of participants and technological setups. Data resulting from this categorization process can be found in the supplementary material for this paper. These data, along with the research knowledge and clinical experience of our extended team and the additional context from the prosthetics literature, ultimately informed the outcomes of this process. 

\subsection{Emerging Themes} 
Adopting EMG-based control systems is inherently complex due to the stochastic EMG signal and subtle changes in user behaviours and environments, so special attention is needed when applying them to novel use cases --- such as those emerging in HCI. While researchers can lean on the work done by the prosthetics community, HCI has distinct challenges in successfully adopting EMG as an input modality. As a result, taking the successes from prosthesis control and applying them directly in HCI for different use cases has been challenging. We briefly expand, below, on each of the four themes that emerged from this process.

\subsubsection{Interaction Design} \label{sec:InteractionDesign}
Humans regularly use their hands and arms to communicate and express themselves in their everyday lives. Extending their use as an intuitive input for interaction between humans and computers is therefore a logical progression. As a result, many HCI researchers have attempted to leverage various input modalities, including EMG-based inputs, to enable these interactions. Despite this, we found a reoccurring theme of insufficient focus on interaction design for use in EMG-based interfaces. One of the main limitations in this regard is the insufficient consideration of the physiological properties of the EMG signal. At the core of any EMG-based interaction is some form of muscular activation that can be categorized (or mapped) and converted into a control input. Researchers must select muscle sites from which appropriate and adequately distinguishable EMG activity can be recorded in response to specific contractions or the interaction will be fundamentally flawed. For example, if electrodes are placed around the forearm, certain finger-based inputs (such as movements of the thumb activated by the thenar eminence muscles, which are intrinsic to the hand) may not generate sufficient EMG activity for robust input detection. It is also imperative for researchers to select muscle-based inputs that coincide with the desired interaction. For example, discrete event inputs (e.g., a button click) should ideally be associated with \textbf{dynamic gesture-based inputs} (e.g., a finger tap) (see section \ref{sec:InputType}) that have a definitive start and end. Many of the surveyed papers, however, naïvely designed interactions based on control scheme precedents set by the prosthetics literature without sufficiently considering the design implications associated with different interactions.

Currently, the predominant use case for EMG centers around enabling amputees to control their prostheses. The interaction between the amputee and the computer during prosthesis control is particular as it requires coordinated and continuous inputs. This \textbf{continuous control} (see section \ref{sec:ControlScheme}) enables amputees to continuously make micro-adjustments to the positioning of their prosthesis (e.g., close the hand a bit more). Enabling this control requires that users sustain predominantly static contractions (see section \ref{sec:InputType}) since input detection is constantly re-occurring. Physiologically appropriate contractions --- those where the user contracts in a manner consistent with the prosthetic degree of freedom they are controlling --- are commonly adopted with pattern recognition-based myoelectric control to improve intuitiveness (e.g., think ``squeeze your hand'' to close the prosthetic terminal device). While this may work for some general-purpose applications like telerobotics, it may be unintuitive and not ideal for others. For example, in the case of smart garments \cite{benatti_towards_2014}, hand-open/close and pinch grip may not be the best inputs as these contractions are commonly elicited during everyday activities and will inevitably trigger false activations. In HCI research we must also consider the intuitiveness of the chosen mappings from input contractions to interaction commands. For example, in the case of drone control \cite{stoica_remote_2014}, we must consider whether wrist flexion and extension are intuitive mappings for forward and backward movement. Additionally, HCI leverages many discrete events as inputs --- those with a definitive start and end --- like clicks, taps, and finger swipes. Unlike prosthesis control, these inputs are not continuous and they are often intended as time-limited event triggers. These command-type inputs are often used as an infrequent supplement to other input modes instead of a dedicated ``always on'' continuous control. For example, gestures could be used to quickly change the song playing through an individual's headphones as they are exercising. In this scenario, continuously recognizing muscular activity would be prone to frequent inadvertent commands, so the interaction design should move from questioning ``Which of the \textit{N} known classes was that? (i.e., continuous control)'' to ``Did a particular event just occur? (i.e., discrete control)''. 

Despite acknowledging this need for discrete event-based inputs, many have endeavored to adopt continuous input schemes to recognize discrete tasks. While this can be made possible, it is an awkward combination, arguably resulting from an over-reliance on the prosthetics literature. For example, consider the design of an EMG-based interaction using finger flexion to activate a button press. Using continuous control, the decision stream would likely output a sequence of ``rest'' (or inactive) decisions, followed by some number of finger flexion decisions (with length governed by the classifier update rate and the speed and length of the finger contraction), followed again by more rest decisions. This decision stream would then have to be mapped to the discrete event through some sequence of filtering or state logic. For example, is the button press registered when the EMG returns to rest, after a fixed number of finger flexion decisions, or after a certain amount of time has elapsed? Additionally, how many consecutive active decisions must occur to register a press (to avoid physiologically impossible activations, such as with switch debouncing in electrical circuits)? Furthermore, the rejection of any contraction other than one of the \textit{N} trained to use the system (which is inevitable during everyday use) can be problematic \cite{robertson_effects_2019}. Although some HCI researchers have recognized the benefit of leveraging inherently discrete gesture commands as inputs \cite{eghtebas_investigation_2018, huang_leveraging_2015, paudyal_sceptre_2016, zhang_hand_2009}, many more have defaulted to using the continuous control approach. For example, to recognize different finger taps in previous work \cite{saponas_demonstrating_2008}, the authors opted for a continuous approach by splitting each input into a series of windows and making a prediction for each window. By leveraging no-movement thresholding and applying a majority vote for the active windows of data, the specific discrete finger tap was selected. With this approach, the temporal structure/envelope of the EMG signal is not fully leveraged, as each finger tap is treated as a subset of individual and unrelated static contractions. Ultimately, this could have negative impacts on the recognition capabilities of this control scheme during real-time use.

\subsubsection{Model Design}
After selecting the appropriate interaction, a control scheme must be developed to accurately and reliably recognize the selected muscle-based inputs. The selection and implementation of the correct control system are imperative to the overall design process and to user experience. An incorrectly designed system will inevitably result in poor recognition, and as a result, the user experience will suffer. We found that many works proposed or used control schemes that heavily borrowed from the prosthetics literature without sufficient consideration for differences in model design and hyper-parameter requirements.

In addition to the widespread adoption of the \textbf{continuous classification} framework by the prosthetics community, its implementation has been heavily guided by its use case. Decades of research have explored the impact of different design parameters, largely building on the early work of Englehart et al. \cite{englehart_robust_2003}. These design considerations include, but are not limited to, the appropriate preprocessing of the EMG signal, the selection of discriminative \textbf{features} \cite{phinyomark_feature_2012, phinyomark_emg_2013}, specific ranges of \textbf{window/increment lengths} to ensure the optimal controller delay \cite{smith_determining_2011}, the selection of appropriate classification schemes, the number and placement of EMG sensors \cite{hargrove_comparison_2007, fang_robust_2014}, and the use of physiologically appropriate contractions as inputs. These parameters have been carefully iterated over time to be optimized for the particular use case of prosthesis control. Although much can be learned from these findings, naïvely applying them to other use cases in HCI may result in sub-optimal control systems. For example, due to the often unique residual musculature of each amputee, these systems have evolved as being \textbf{user-dependent} (with a new model trained for each new user), with only recent consideration of cross-user models \cite{campbell_deep_2021}. While these cross-user models are key to the future success of EMG-based input in HCI, they were of little focus in the sampled work. Furthermore, the assumed stationarity of EMG within the short windows used in the continuous control paradigm has largely trivialized the use of temporally aware classification schemes \cite{chan_continuous_2005}. For general HCI, however, the selection of interaction design may result in substantially different control scheme requirements. Consequently, although a valuable starting point, HCI researchers should be cautious and discerning in how they leverage knowledge from the prosthetics field for use in HCI.

\subsubsection{System Evaluation}
After selecting, designing, and implementing an EMG-based control scheme, it must be evaluated to determine how well it performs compared to previous works. For EMG-based control schemes, there are generally two agreed-upon forms of evaluation: \textbf{offline evaluation} and \textbf{online evaluation} (see section \ref{sec:Evaluation}). Offline evaluation refers to the performance assessment of algorithms on a pre-recorded set of data. Although this can be preferable when comparing many different control schemes or hyper-parameters, the user is not part of the control loop and therefore, cannot correct for errors. As a result, metrics such as classification accuracy often correlate poorly with the online usability of the resulting control systems \cite{lock_real-time_2005, hargrove_real-time_2007, nawfel_multi-variate_2021}. The substantial challenges associated with measuring online usability for prosthesis research, such as expensive prosthesis fittings and accessing amputee participants has led to the widespread use of such offline metrics. These metrics have translated into HCI research as well, where approximately half of the sampled papers focused solely on offline metrics (i.e., classification accuracy) as their only evaluation criteria. This reliance stems from the previous notion that ``accuracy is undoubtedly one of the fundamental performance metrics for any recognition system'' \cite{paudyal_sceptre_2016}. The focus on offline over online evaluations aligns with classic tension between evaluations that focus on internal over external and ecological validity \cite{mackenzie2012human}. However, a focus on offline evaluation, which overly prioritizes internal validity, provides little insight into how successful a control scheme would be in real-world use. Previous work has established that data recorded in offline settings differ from those seen during real-time use \cite{chang_wearable_2020}. Therefore, user studies should be designed to maximize external and ecological validity. 

Similarly, the challenges with recruiting amputees may have normalized the use of small numbers of participants in EMG research. HCI EMG research has tended towards smaller studies with very few participants (often less than 10), with some testing on as few as one pilot participant \cite{asai_finger_2019, li_automatic_2010, tortora_dual-myo_2019}. The issues around the lack of external and ecological validity and small sample sizes in these studies drastically limit the conclusions that can be drawn from previous work. While, arguably, HCI EMG work is still new, evaluations need to mature to the standards expected of the broader HCI community. 

\subsubsection{Reproducibility}
To grow and evolve the field of EMG-based control in HCI, we must be able to reproduce the work done by others. This is especially true in HCI, where reproducibility has become a popular topic in recent years \cite{wilson_replichi_2011,echtler_open_2018}. Reproducibility, however, is especially complicated when dealing with EMG and has been a struggle for both prosthesis and HCI researchers alike due to the unique EMG patterns among participants and confounded by different technologies and evaluation protocols. While some have called for better standardization \cite{phinyomark_emg_2018}, we found that many studies used inconsistent terminologies, failed to define or quantify parameters, used custom hardware, and lacked sufficient detail for replication. This ultimately hinders the impact of such works, and correspondingly, the growth of the field. Furthermore, whether due to research ethics restrictions, to maintain control over one's own research, or otherwise, many datasets and source code have not been made publicly available. Without access to, and sufficient context for these data and code, it is often impossible to discern why a control system may have achieved the performance it did. We correspondingly suggest that all datasets, hardware specifications, source code, and anonymized participant information should be published alongside the associated work to further encourage collaboration and validation. The public Nina Pro dataset \cite{atzori_building_2012} is a strong example to build on and includes all of these elements. Additionally, EMBody \cite{karolus_embody_2021} is another strong example that released all source code and hardware schematics. Making work publicly available will lead to better accessibility, validity, and ultimately research in this space. In general, we believe that the adoption of standardized approaches to EMG-based HCI research could help to propel the field forward. 

\begin{figure*}[h!]
  \centering
  \includegraphics[width=16cm]{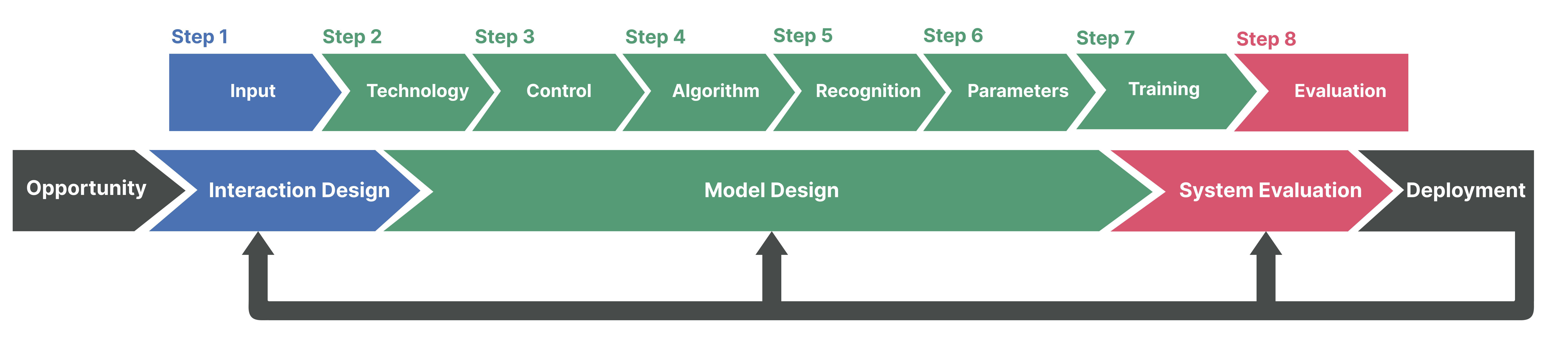}
  \caption{Each individual step of the proposed framework for designing EMG based control systems.}
  \Description{The linear progression of the proposed framework is displayed with the main steps (i.e., Interaction Design, Model Design, and System Evaluation) and their corresponding steps above. We also display an arrow at the end of the linear progression to indicate that this process is somewhat cyclic in nature.}
  \label{fig:ProposedFramework}
\end{figure*}

\section{A Framework for Building EMG-Based Interactions} \label{sec:section4}
To facilitate the design of EMG-based control systems, we propose a design framework (outlined in Figure ~\ref{fig:ProposedFramework} and summarized in Table ~\ref{table:Overview}) consisting of three main categories --- each aligning with one of the extracted themes above: (1) Interaction Design, (2) Model Design, and (3) System Evaluation. The fourth theme of reproducibility is addressed indirectly through the creation of this framework. The framework begins with an interaction opportunity and follows eight steps along the path to deployment, with connections to promote iteration, as is common for robust design processes. This framework was refined through conversations with members of our larger research team (including experienced HCI and prosthetics researchers). It is important to emphasize that this framework is built on best practices and does not propose new steps. Instead, it is the first time, to the best of our knowledge, that these steps have been documented and organized in one place with a focus on EMG-based HCI research. 

\subsection{Interaction Design}

\subsubsection{Step 1: Input Type} \label{sec:InputType}
\begin{figure}[h!]
  \centering
  \includegraphics[width=\linewidth]{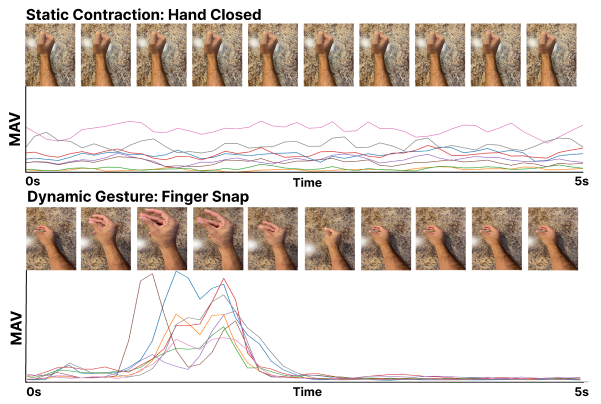}
  \caption{The mean absolute value (MAV) of two separate EMG recordings --- a static contraction (hand closed) and a dynamic gesture (finger snap). In this example, the MAV is a representation of the EMG envelope over a five-second period. Each line of the graph corresponds to an individual channel of EMG data.}
  \Description{We display two different time plots. The first-time plot shows the MAV over a five-second recording of the hand-closed contraction. The MAV of this plot is relatively linear with minor fluctuations over time. On top of this plot are photos of the hand-closed contraction to indicate the corresponding muscular input for each moment in time. The second time plot displays the MAV over a five-second recording for the finger snap gesture. Similarly, a photo corresponding to each gestural component is above the time plot. In this plot, there is clearly a variation in the MAV as the individual transitions through the gesture.}
  \label{fig:ContractionsAndGestures}
\end{figure}

The initial step when designing any EMG-based interaction is to consider the muscle-based input that will enable the desired interaction. Ideally, these inputs should reflect the context of their eventual use (including situational, environmental, and other task-related contexts), as this will affect their efficacy for a given interaction. The literature contains a plethora of terminologies used to describe muscle-based inputs such as poses, contractions, static, and dynamic gestures. To eliminate confusion, we categorize all EMG-based inputs into two categories: \textbf{static contractions} and \textbf{dynamic gestures}. Static contractions are sustained over a period of time, and thus have no meaningful temporal change in the EMG envelope (i.e., a sequence of EMG activity) over that period (e.g., squeeze your hand closed). Conversely, dynamic gestures inherently involve a changing pattern of EMG consistent with the associated gesture (e.g., a finger snap). Figure ~\ref{fig:ContractionsAndGestures} below highlights the differences between the two.

Because static EMG contractions are sustained and relatively constant, there is little added benefit in analyzing their temporal envelopes (i.e., how the EMG evolves over time). This enables the extraction of features from short windows in which the data are assumed to be stationary (see section \ref{sec:Parameters}). As a result, static contraction recognition can occur at any point during elicitation such as in continuous prosthesis control (i.e., ``always on'' repeated classifications). Another added benefit of static contractions is that they often require less computationally complex algorithms and may be easier to recognize due to their reduced variability as compared to dynamic gestures. A caveat, however, is that even though the EMG patterns are considered static, variations in the intensity of the contraction have been used to enable proportional control over the velocity of prosthetic devices \cite{campbell_current_2020, fougner_control_2012,scheme_motion_2014}.  
In EMG-based control, the concept of a \textit{static contraction} must be differentiated from the term \textit{posture}, which implies a given position of the limb. While a useful kinematic input for modalities such as computer vision, maintaining a posture does not necessarily require the contraction of muscles to generate EMG. For example, depending on gravity, a user's hand could be resting in a hand-open posture without eliciting a corresponding hand-open EMG contraction. Conversely, their hand could be constrained in a closed posture (e.g., tucked in a pocket), but the user could be intentionally activating a ``finger extension'' EMG contraction. This difference between contraction and position can confound novice EMG-based interaction users without sufficient training but presents interesting design opportunities for the use of EMG in HCI.

We define dynamic gestures as a sequence of contractions elicited as part of a discrete event (e.g., a finger snap). This implies that dynamic gestures comprise at least a beginning and end state, with an evolving sequence of EMG in between. As such, gestures have a temporal envelope that must be considered for accurate recognition, allowing for natural variations in EMG and completion rates. Because of this temporal structure, the recognition of gestures typically occurs after the completion of the gesture. While there are techniques for predicting gestures midway through completion \cite{mori_early_2006,izuta_early_2014}, waiting until the end is usually the most robust detection method. For many HCI applications, gesture-based inputs may arguably be more natural (i.e., more similar to the hand/arm movements we use to communicate) and map intuitively to the control of discrete outputs. Potential examples of dynamic gestures include some sign language words, wrist flicks, finger swipes, and finger snaps. 

\subsection{Model Design}
After selecting the appropriate interaction, a control system must be designed to reliably recognize the desired EMG-based input. In prosthesis control, the model design is an important part of the broader design, but factors such as the socket fit, device mechanics, component weight, and hardware limitations also play significant roles in overall performance. For general-purpose HCI applications, these same factors may not apply, potentially increasing the relative impact of model design on the overall system. We categorize the main considerations for model design into six steps: Technology Selection, Control Scheme, Algorithm Selection, Recognition Type, Parameter Selection, and Model Training.

\subsubsection{Step 2: Input Technology Selection} \label{sec:TechSelection}
When designing any EMG-based input system, the appropriate input technology must be selected to enable the desired interaction. Incorrect technology selection from the beginning will inevitably hinder the ability of a system to recognize inputs reliably. Four factors that should be considered during this stage include (1) Sampling Rate, (2) Number of Electrodes, (3) Electrode Location, and (4) Multi-Sensor Fusion (such as with the inclusion of an Inertial Measurement Unit (IMU)). These factors are summarized in Table ~\ref{table:TechnologySelection}. 

\begin{table*}[htb]
  \caption{Factors Influencing Technology Selection}
  \begin{tabular}{p{0.16\linewidth} | p{0.79\linewidth}}
    \toprule
    \textbf{Consideration}  & \textbf{Description} \\
    \midrule
    Sampling Rate & The sampling rate of an sEMG device --- measured in hertz (Hz) --- is the number of times per second that a reading is taken from the electrodes. Sampling theory suggests that EMG should be sampled at above 1000 Hz, however, pattern recognition performance plateaus around 500 Hz \cite{phinyomark_feature_2018}. Note that most EMG energy is between 50-150 Hz, with the usable energy being between 0-500 Hz \cite{phinyomark_emg_2018}. \\
    \midrule
    Number of Electrodes & Increasing the number of electrodes can improve recognition through better spatial distribution of the EMG sensors, but performance is dependent on placement location and target gestures/contractions. For cuffs mounted circumferentially around the forearm, diminishing returns have been found above 6 to 8 electrodes \cite{hargrove_comparison_2007}.  \\
    \midrule
    Electrode Location & The location on the arm where the electrodes will be placed (e.g., forearm or wrist) affects which inputs can be accurately detected \cite{botros_electromyography-based_2022} and the type of electrodes that can be used (e.g., electrode cuffs or individual electrodes). \\
    \midrule
    Sensor Fusion & The inclusion of other sensing modalities can be beneficial for providing complementary information to EMG. For example, the use of an IMU can provide gross arm movement information not provided by wrist or forearm mounted EMG \cite{zhang_hand_2009, scheme_examining_2010}. As we move toward recognizing muscular inputs during everyday activities, these additional sensors and optimizing their use and even inclusion based on situational context may eventually become a core component of the framework. Currently, however, sensor fusion is an ongoing area of research. \\
  \bottomrule
  \end{tabular}
  \label{table:TechnologySelection}
\end{table*}

\bigbreak
 
\subsubsection{Step 3: Control Scheme} \label{sec:ControlScheme}
\begin{figure}[h]
  \centering
  \includegraphics[width=\linewidth]{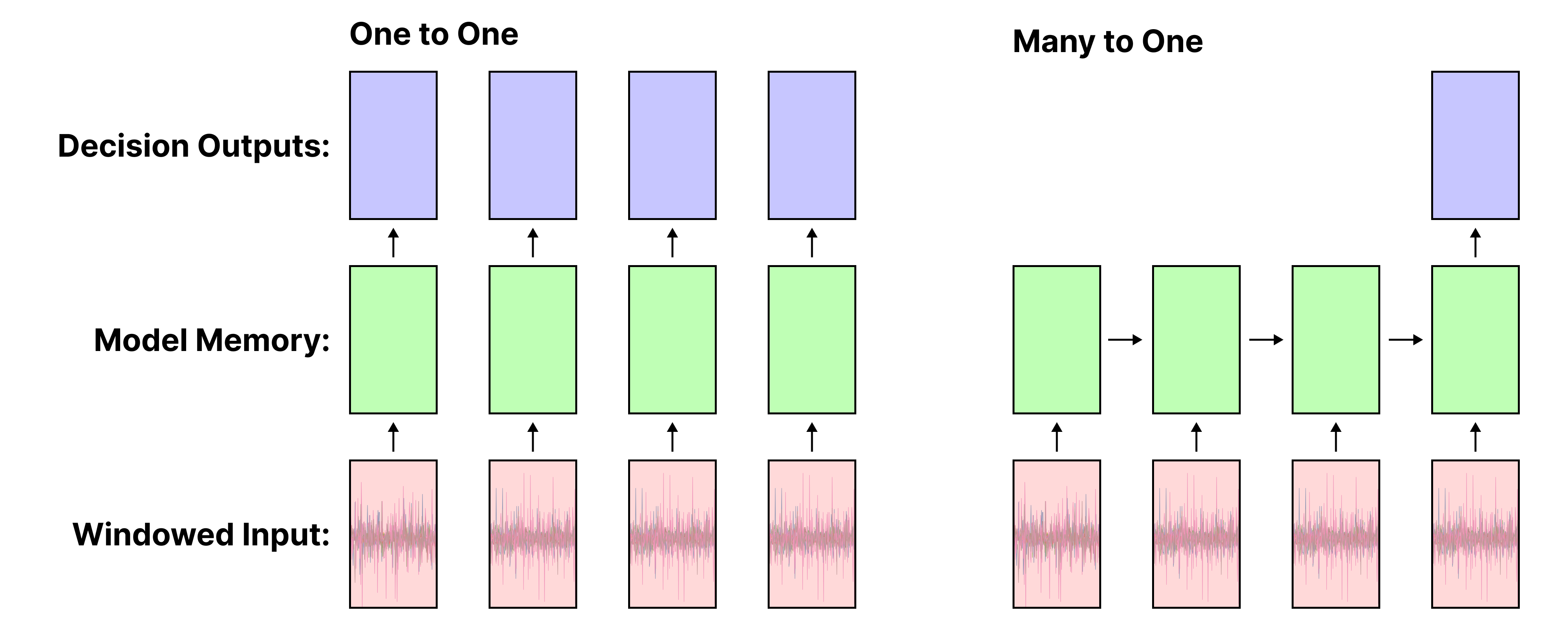}
  \caption{The two main control scheme mappings from windowed EMG inputs to output decisions; one-to-one mappings for continuous control outputs (left), and many-to-one mappings for the discrete recognition of dynamic gestures (right).}
  \Description{We display two different mapping structures (one-to-one) and (many-to-one) using three levels of blocks. The first level is a set of four blocks representing the windowed EMG input. For both mapping structures there are arrows pointing up to the next set of four blocks representing the model memory. For the one-to-one mapping there are no arrows between the model memory blocks. In the many-to-one mapping there are arrows between the blocks to indicate that the model is passing along the information. Finally, the one-to-one mapping outputs a decision for every input, whereas the many-to-one mapping only outputs one decision at the end.}
  \label{fig:MappingStructure}
\end{figure}

A fundamental aspect of any EMG-based control scheme is the set of predefined static contractions or dynamic gestures that generate an event. These events are based on the underlying EMG and occur after a model has made an output decision. Once this happens, the event is processed by a controller and converted into a control input. How and when these events are generated drastically influences the behaviour of the control scheme, particularly whether they are irregular (based on discrete inputs) or continuous (based on a predefined fixed interval). As such, we split the event generation process for EMG control into two distinct control schemes: \textbf{continuous control} and \textbf{discrete control}. Figure \ref{fig:MappingStructure} exemplifies the difference in mapping structures between the two.

A continuous control scheme generates a sequence of regular periodic events based on a defined parameter (e.g., window length and increment). These events are continuously passed to the application and converted to a device function, and computed from a single short window of EMG data, in which the data are assumed to be stationary. In this way, the continuous control scheme can be largely considered as a one-to-one mapping (one window of EMG to one decision event). Applications leveraging this control scheme use static contractions as inputs and require a focused effort from the user due to the repeated mapping of EMG to control input. While this is useful when an application requires a continuous stream of events (e.g., controlling the movement of a cursor or a prosthetic limb), it is not universally advisable for many HCI applications.

The discrete control approach is ideal for dynamic gesture-based inputs as it generates a single event after the completion of a pre-defined set of actions with a definitive start and end. A temporal model is used to learn patterns in the sequence of transitions between internal states (contractions) and generate a particular event when such a pattern is recognized. Because these systems only respond after the completion of a sequence of states (a dynamic gesture) the outputs are discrete and sporadic and can be considered as employing a many-to-one mapping (many windows of EMG to one decision event). This imposition of temporal structure can improve the robustness of EMG pattern recognition to spurious activations and require less focused effort on the part of the user (inputs are only activated when a specific condition is met).

\subsubsection{Step 4: Algorithm Selection} \label{sec:AlgSelection}
To enable multi-channel EMG-based control systems, machine learning models learn to discriminate between patterns of muscular inputs. Prosthetics researchers have tested almost every available machine learning algorithm to classify EMG activity with mixed results \cite{chan_continuous_2005, campbell_linear_2019, oskoei_support_2008, cote-allard_interpreting_2020}. Therefore, algorithm selection for a new system can be daunting because there is no perfect or superior candidate for all use cases. To simplify the process, we group popular machine learning models into two categories: \textbf{stationary models} and \textbf{temporal models}. Selecting the category of algorithms useful for the designed interaction is important for reliable recognition.

The windowing of static contractions in continuous control lends itself to classification using stationary models. Because these windows are short, the contractions are assumed to be constant, alleviating the need for treatment of any temporal structure in the EMG envelope. Examples of stationary models are Linear Discriminant Analysis (LDA) \cite{campbell_linear_2019}, Quadratic Discriminant Analysis (QDA) \cite{ghojogh_linear_2019}, Support Vector Machines (SVM) \cite{chang_libsvm_2011}, K-Nearest Neighbour (KNN) \cite{jiang_survey_2007}, Artificial Neural Networks (ANN) \cite{jain_artificial_1996}, Convolutional Neural Networks (CNN) \cite{zhai_self-recalibrating_2017,lecun_gradient-based_1998}, and Random Forest \cite{biau_random_2016}.

The temporal structure inherent in repeatable dynamic gestures is best recognized using temporal models. Temporal models, therefore, rely on more than one window of data to understand the evolution of the EMG signal over time, and may require the optimization of additional hyperparameters related to the dynamics of the EMG gesture. Examples of temporal models include Long Short-Term Memory (LSTM) models \cite{greff_lstm_2017,jabbari_emg-based_2020}, Temporal Convolutional Networks (TCN) \cite{bai_empirical_2018, zanghieri_robust_2020}, Hidden Markov Models (HMM) \cite{zhang_hand_2009, hu_comprehensive_2020}, and Dynamic Time Warping (DTW) \cite{berndt_using_1994, huang_emg-based_2010}.


\subsubsection{Step 5: Recognition} \label{sec:Recognition}
In classification-based machine learning each input/interaction recognized by the system is called a class. The goal of the machine learning model is then to match an input to its correct class, assuming that the patterns of EMG are repeatable and separable. In prosthesis control and in HCI, subtle variations in how users elicit contractions \cite{tabor_designing_2017} and differences introduced by confounding factors such as electrode shift or limb position \cite{campbell_current_2020} have been shown to reduce the repeatability of patterns. Training with data that represent all possible situations that may occur during online use can improve the robustness of recognition models \cite{scheme_electromyogram_2011} but is tedious and unrealistic for many use cases. Recognition that different and potentially unwanted patterns may be generated gives way to the consideration of \textbf{closed-set} and \textbf{open-set} recognition.

In closed-set systems, a classifier always outputs a decision consisting of one of a predefined set of class labels (i.e., 1 of \textit{N} classes) with the assumption that the user will only generate contractions consistent with one of the trained classes. While these models have a limited worldview, they are simpler to implement since the possibility of external classes does not need to be considered. However, this also means that these systems are prone to false activations, particularly if users are distracted by other tasks. While this form of recognition is used heavily in ``always on'' continuous control, it falls apart when users elicit contractions outside of the closed set of classes. Some forms of post-processing, such as the rejection of low confidence decisions to an inactive class, can alleviate some of these limitations \cite{scheme_confidence-based_2013}.

Conversely, open-set recognition considers the possibility that the \textit{N} known classes are only a subset of a larger set of possible actions. As a result, these models must be aware that current contractions or gestures may not belong to one of the trained classes, giving them a more robust worldview. The implementation of such models, however, can be more complicated as it is difficult to differentiate between a small target set of contractions/gestures and all possible (and therefore unknown) EMG patterns. There has correspondingly been relatively little work in open-set recognition of EMG, however, its solution could yield more robust models for general use in HCI \cite{chang_wearable_2020}.

\subsubsection{Step 6: Control Parameter Selection} \label{sec:Parameters}
\begin{figure}[h]
  \centering
  \includegraphics[width=\linewidth]{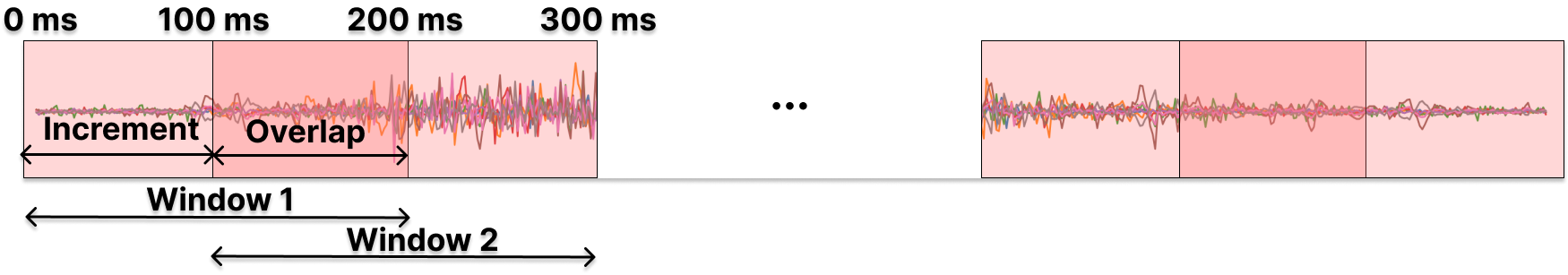}
  \caption{Windowing of raw EMG data.}
  \Description{The windowing process is displayed on a segment of raw EMG data. Generally, the segment is shaded with overlapping boxes to highlight windows, increments and overlaps.}
  \label{fig:Windowing}
\end{figure}

The selection of appropriate parameters is critical to the performance and robustness of any control system. An important parameter in EMG-based control is the window length and increment. Because EMG is stochastic, features must be extracted that describe the signal characteristics. To be usable in a control interface, these features must be computed from a relatively short segment of EMG data (150-250 ms in prosthesis control \cite{farrell_optimal_2007}). Using longer windows can improve recognition rates through more robust feature estimation but they can also decrease controller responsiveness as decisions are influenced by older data. The window increment dictates how much the algorithm steps forward in time before extracting a new window, and thus how frequently classification decisions are computed (see Figure ~\ref{fig:Windowing}). Shorter increments result in more responsive systems (and a more dense event decision stream in continuous control) but higher processing requirements. Balancing the tradeoff between controller delay, responsiveness, and computational requirements is important for researchers to consider and justify when designing EMG-based control systems. 

The extraction of features from EMG windows serves to increase the information density before classification. For decades, the prosthetics community has explored and evaluated new feature sets to improve prosthesis control. Phinyomark et al. have contributed robust guides on feature selection by exploring and explaining ideal features for prosthesis control \cite{phinyomark_feature_2012, phinyomark_navigating_2017} and lower sampling rate EMG wearables \cite{phinyomark_feature_2018-1, phinyomark_feature_2018-2}. Many other studies have also explored the role of features and reducing feature redundancy \cite{hudgins_new_1993, phinyomark_novel_2009, javaid_comparative_2018}. These validated feature sets should be leveraged by researchers based on the selected input devices and chosen control parameters. 

It is important to note that many of these parameters have been optimized by the prosthetics community for their specific use case. Therefore, while these parameters should work in general-purpose applications that employ continuous control, additional consideration may be required when implementing discrete control for gesture recognition. For example, delays induced by slightly longer windows may not be as noticeable in the context of a longer dynamic gesture, or perhaps shorter frame increments may improve recognition of temporal patterns. Furthermore, it is not yet clear whether the features identified for continuous control will capture, or be robust to, the dynamics required for discrete inputs. Understanding and exploring these intricacies of parameter selection may therefore be important for the continued advancement of discrete control. 

\subsubsection{Step 7: Model Training}\label{sec:ModelTraining}
Conventional model training involves recording representative EMG data for each different input contraction or gesture and then training the model to discriminate between the various classes. Several considerations must be made before training, including a determination of the acceptable user- and context-dependence of the model. The majority of EMG models in the literature and in practice have been \textbf{user-dependent}, meaning that they are trained using one user's data and are thus specific to that user. This is required in prosthetics due to differences in the residual musculature in amputees, but also remains an outstanding challenge for able-bodied EMG users \cite{Campbell_Chang_Phinyomark_Scheme_2020}. The quality of the training data can also greatly affect the \textbf{context-dependence} of a model even when user dependent. The inclusion of a variety of factors such as arm position, contraction intensity, and electrode positioning has been found to reduce the impact of these common confounding factors at the cost of increased training time. Continued research in \textbf{user-independent} models \cite{campbell_deep_2021}, particularly for general-purpose HCI applications, could therefore improve the end-user experience by reducing the training time required and facilitate \textbf{context-independence} by leveraging pre-recorded datasets from other users.

\subsection{System Evaluation}
After the successful design and implementation of an EMG-based control system, the final step should always be system evaluation. The goal of this stage is to understand how the newly designed system performs and compares to previous works. Although there are many disparate methods for evaluation, they can largely be grouped into \textbf{offline} or \textbf{online evaluation} techniques. Both techniques are important tools when evaluating the viability and overall impact of new control systems, but each has its limitations. 

\subsubsection{Step 8: Evaluation} \label{sec:Evaluation}
The majority of EMG-based control system literature has used offline analysis techniques to evaluate the performance of control schemes. Classification accuracy (the percentage of correct output decisions) has been widely adopted in continuous control research but while it gives some insight into a system's performance, it does not fully explain online usability \cite{lock_real-time_2005, nawfel_multi-variate_2021}. In discrete control, the temporal nature and relative infrequency of gesture-based inputs may also contribute to the inadequacy of accuracy as a metric, and thus additional metrics such as false negatives (missed gestures), false positives (accidental activations), and response time (the time between gesture initiation and recognition) should be considered. Nevertheless, while offline metrics are useful when developing initial models --- in part because the same data can be used to evaluate many different models --- they are not a replacement for evaluating online usability. As such, HCI researchers developing EMG-based interfaces should focus on online evaluation techniques whenever possible.

Online evaluation focuses on user ``in-the-loop'' interactions to simultaneously assess the performance and usability of an EMG-based control system. Metrics may vary by application but should include controlled objective comparisons against state-of-the-art and baseline systems such as recognition rates during real-time use \cite{delpreto_plug-and-play_2020} or throughput as computed using Fitts' Law frameworks \cite{scheme_validation_2013, tabor_quantifying_2017}. Questionnaires (e.g., NASA TLX or SUS) \cite{kerber_same-side_2015} and qualitative responses via interviews \cite{karolus_hit_2020} can provide important and complementary added context. Although the actual evaluation tasks are dependent on the interaction and control system being tested, the user should be allowed to interact with the designed control system. Their interaction with the control scheme is critical not only for overall performance evaluation but in understanding how the control scheme responds to natural variations in user inputs and their responses to recognition errors. Given our focus as a community on the interaction between humans and computers, offline analyses should be used sparingly, and online evaluation should be strongly prioritized. 

\subsection{Example Applications of the Framework}
\label{framework-examples}
To exemplify how this framework can guide the design of EMG-based input, in this section, we provide a brief walkthrough of its application in the design of two EMG-based control systems. We have selected two examples from previous work. The first system controls a remote-control (RC) car \cite{kim_emg-based_2008} in four directions (up, down, left, and right) (see Table \ref{table:Example1}). The second is for music control during exercise \cite{saponas_enabling_2009} (see Table \ref{table:Example2}). The goal of this system is to enable the hands-free control of a music application (i.e., volume control, play/pause control, and song skipping) while jogging. These examples were selected as they follow distinct paths in our framework. 

\begin{table*}
  \caption{RC Car Controller Example}
  \begin{tabular}{p{0.2\linewidth} | p{0.75\linewidth}}
    \toprule
    \textbf{Step} & \textbf{Choice and Reasoning} \\
    \midrule
    Input Type & \textbf{Static contractions} are selected to enable the continuous control of direction commands. Based on a right-handed user, control mappings might include: wrist flexion (Left), wrist extension (Right), hand close (Down), and hand open (Up). \\
    \midrule 
    Technology & An \textbf{inexpensive armband}, sampling at 200-500 Hz would likely be sufficient, as this captures most of the usable EMG energy and is therefore adequate for this relatively simple five class problem. For the chosen inputs, the band would probably be placed around the participant's forearm. No IMU inputs are required for recognizing the chosen controls. \\
    \midrule 
    Control Scheme & \textbf{Continuous control} is chosen because RC control requires the constant adjustment of the direction of the vehicle. \\
    \midrule
    Algorithm & A \textbf{stationary model}, such as an LDA, could be selected based on the use of static contractions and due to its robustness and simplicity. \\
    \midrule 
    Recognition & \textbf{Closed-set recognition} could be selected, assuming that the user is focused and unlikely to be performing other activities while controlling the RC car. \\
    \midrule 
    Parameter Selection & Because this system requires regular and fast response times, using literature norms of a \textbf{window size of 200 ms} and an \textbf{increment of 50 ms} might be a good place to start. \textbf{Phinyomark's TD4 feature set} \cite{phinyomark_feature_2018}, which is optimized for wearable devices, could be adopted.\\
    \midrule 
    Model Training & For better performance, \textbf{user and context-dependent} models could be trained, unless convenience is a priority. Additionally, the collection of variable training data \cite{scheme_training_2013} could improve robustness. \\
    \midrule 
    Evaluation & \textbf{Online evaluation} should be adopted, as the goal is the usable control of the vehicle. An example of this would be to compare the user's ability to drive the car through an obstacle course using the EMG-based control system versus an alternate controller (e.g., joystick). \\
    \bottomrule
  \end{tabular}
  \label{table:Example1}
\end{table*}

\begin{table*}
  \caption{Music Control While Exercising Example}
  \begin{tabular}{p{0.2\linewidth} | p{0.75\linewidth}}
    \toprule
    \textbf{Step} & \textbf{Choice and Reasoning} \\
    \midrule
    Input Type & The interactions with a music interface (e.g., pressing a pause button) are infrequent and discrete, motivating the use of discrete \textbf{dynamic gesture}-based inputs (i.e., a button click). Examples of mappings include hand wave (start/stop), wrist flick up/down (volume up/down) and double tap (next song). \\
    \midrule 
    Technology & To facilitate use during exercising, the technology should be low profile, ideally in the form of a \textbf{wearable} and have a sufficient sampling rate to capture dynamics for reliable recognition (e.g., 500-1000 Hz). \\
    \midrule 
    Control Scheme & \textbf{Discrete control} enables the recognition of discrete actions for controlling the music player. \\
    \midrule
    Algorithm &  A \textbf{temporal model} should be selected to recognize the temporal patterns of the dynamic gestures. \\
    \midrule 
    Recognition & \textbf{Open set recognition} is crucial for this use case because the user is actively engaged in other, non-target, muscular activities. A unique wake word could be adopted to improve robustness to false activations (see section \ref{sec:FalseActivations}). \\
    \midrule 
    Parameter Selection & Given its complexity and relative novelty, further \textbf{experimentation} would be required to determine optimal model parameters. \\
    \midrule 
    Model Training & The demanding and changing environment of physical exercise, and the potential for electrode shift, sweat, and muscle fatigue, suggest that the model should be \textbf{context-independent}.  \\
    \midrule 
    Evaluation & Although initial model validation could be done \textbf{offline}, robustness should be tested in an \textbf{online} setting where users interact with the control system while exercising. \\
    \bottomrule
  \end{tabular}
  \label{table:Example2}
\end{table*}

\begin{table*}[h!] 
  \caption{Framework Summary}
  \label{tab:freq}
  \begin{tabular}{p{0.2\linewidth} | p{0.2\linewidth} | p{0.55\linewidth}}
    \toprule
    \textbf{Category} & \textbf{Subcategory} & \textbf{Description} \\
    \midrule
    \multicolumn{3}{c}{Interaction Design} \\
    \midrule
    Input Type        & Static Contraction              & A fixed muscle contraction that is sustained (i.e., EMG does not intentionally change over time). \\
                      & Dynamic Gesture             & A muscle contraction that varies dynamically over time (i.e., EMG intentionally changes in a specific way over time).             \\
    \midrule
    \multicolumn{3}{c}{Model Design} \\
    \midrule
    Technology Selection    & Factors to Consider      & Factors to consider include: (1) Sampling Rate, (2) Number of Electrodes, (3) Electrode Location, and (4) Multi-Sensor Fusion.  \\
    
    Control Scheme          & Continuous             & Output events are regular and periodic based on defined parameters such as window length and increment.             \\
                            & Discrete              & Output events are irregular and occur after the completion of a discrete sequence of actions with a definitive start and end.           \\
    Algorithm Selection     & Stationary Models    & Conventional static machine learning models that \textbf{do not} explicitly consider the temporal structure of inputs.  \\
                            & Temporal Models       & Temporal machine learning models that explicitly model the temporal structure of dynamic inputs. \\
    Recognition             & Closed Set             & Outputs are assumed to always belong to the set of \textit{N} known classes (one of which may be a “do nothing” class). \\
                            & Open Set              & Outputs are assumed to be part of a larger set of \textit{M} possible actions, of which the \text{N} known actions are a subset. \\
                            
    Control Parameters     & Windowing                & The process of subdividing the EMG time series into regular, fixed-length segments from which features can be extracted. Windows may be non-overlapping, or overlapping (using some increment smaller than the window length) to increase the temporal resolution of output decisions.              \\
                            & Features                 & Feature extraction increases the information density of the underlying EMG by computing descriptive properties from a window of data. The selection of features used for a control system greatly influences its performance. \\ 
    Model Training          & User Dependence          & \textbf{User dependent} models are specific to a particular user and may not generalize to other users, whereas \textbf{user independent} models work across multiple users.   \\
                            & Context Dependence       & A model that is sensitive to confounding factors known to degrade EMG patterns (such as changes in electrode placement or limb position) is considered \textbf{context dependent}. Models that are robust to such confounding factors are termed \textbf{context independent}.       \\
    \midrule
    \multicolumn{3}{c}{System Evaluation} \\
    \midrule
    Evaluation    & Offline           & Model evaluation is performed using EMG data that were previously recorded without explicit user feedback. Enables the direct comparison of multiple models on the same dataset. \\
                    & Online            & System evaluation is conducted with the user ``in-the-loop'', responding to feedback from the control scheme under test. The resulting data cannot be used for further offline model development, because they are impacted by the user's responses to that control scheme . \\
                  
    \bottomrule
    \end{tabular}
    \label{table:Overview}
\end{table*}

\section{Further Challenges and Considerations} \label{sec:section5}
We acknowledge that the design framework outlined in the previous section does not comprehensively cover all of the intricacies and complexities of EMG-based control. Instead, we aim to provide a simple framework to guide the design thought process and serve as a starting point for researchers getting into this field. In this section, we explore further challenges and considerations, drawing on current concepts and challenges in prosthetics. We group these considerations into five categories: Signal Quality, False Activations, User Training, Adaptation, and Activity Detection. As the field evolves and expands, and with increased research in HCI, so too will this list. 

\begin{table*}[h!]
  \caption{Common Noise Sources to Consider When Acquiring EMG}
  \begin{tabular}{p{0.2\linewidth} | p{0.375\linewidth} | p{0.375\linewidth}}
    \toprule
    \textbf{Type of Noise}  & \textbf{Description} & \textbf{Solution} \\
    \midrule
    Motion Artifact & Low-frequency noise added to the EMG due to the movement of the surface electrodes relative to the skin \cite{mccool_identification_2014}. & Ensuring secure electrode-skin contact and high-pass filtering can help to reduce the impact of this noise \cite{de_luca_filtering_2010}. \\
    \midrule
    Power Line Interference & Band-limited noise added to the EMG due to electromagnetic power line interference on the human body \cite{tomasini_power_2016}. & The simplest method for eliminating power line interference is by using a notch filter at the power line frequency of 50 or 60 Hz \cite{tomasini_power_2016}. \\
    \midrule
    Equipment Noise & As with all electronic equipment, electric noise can arise when measuring signals. & Selecting high-quality surface electrodes can help alleviate equipment noise \cite{raez_techniques_2006}. \\
  \bottomrule
  \end{tabular}
  \label{table:NoiseSources}
\end{table*}

\subsection{Signal Quality}
The quality of the input signal is critical for machine learning models as they rely on repeatable and separable patterns for accurate recognition \cite{chang_assessment_2020}. As an electrophysiological signal, however, EMG is prone to noise corruption from sources such as power line interference and motion artifact (movement of the sensors relative to the skin). Consequently, signal quality has been a priority for the prosthetics community, where electrodes are embedded in sockets that may shift with movement. As a result, there have been many strategies employed to (1) detect signal noise \cite{spanias_detection_2016,tanweer_development_2019} and (2) remove noise from the signal \cite{mewett_removing_2001, de_luca_filtering_2010}. These techniques should similarly be employed in HCI research to ensure the most robust interactions possible. A list of common noise sources and methods to alleviate them can be found in Table ~\ref{table:NoiseSources}.

\subsection{False Activations} \label{sec:FalseActivations}
An undesirable characteristic of most control schemes is the presence of \textbf{false activations}, which occur when the system erroneously outputs an unintended action. In EMG-based control schemes, accidental muscle contractions or those generated during regular movement may result in false activations and can detract from the user experience and embodiment \cite{hargrove_multiple_2010}. This is similar to the ``Midas Touch'' problem experienced with other input modalities such as eye tracking \cite{midas_touch}. In prosthesis control, false activations can lead to catastrophic results such as dropping or crushing objects (e.g., spilling a hot beverage or breaking an egg). In general-purpose HCI applications, false activations may cause unexpected input commands, frustrating the user and detracting from a sense of agency. Because of these implications, extensive research has been done to help alleviate the issues caused by misclassifications.

\textbf{Rejection} \cite{scheme_confidence-based_2013} is a technique wherein classifier outputs are overridden to a default or inactive state when the output decision is unsure. This concept stems from the notion that it is often better (less costly) to incorrectly do nothing than it is to erroneously activate an output. This is especially beneficial in continuous prosthesis control, where the next event is assured to occur momentarily. Over-rejection, however, can be frustrating and leave the user with a sense of `frozen' control. In discrete control systems, this tradeoff is also important because of the relative infrequency of decisions and the potential for user frustration, either from false activations or over-rejection, requiring repeated attempts to activate an output. Nevertheless, rejection is a simple and important tool that can reduce false activations in general-purpose HCI applications.

\textbf{Post Processing} is a group of strategies that leverage additional context, such as from previous outputs, to inform the current output. Majority voting \cite{geng_gesture_2016,wahid_multi-window_2020} is a popular method that overrides the current output with the label corresponding to the class that occurred most frequently over the past \textit{N} decisions. As a form of simple low-pass filter, this introduces a delay into the system but reduces the likelihood of spurious false activations. While majority voting is only practical for continuous control, other forms of post processing could be used for discrete control applications. For one, a form of timeout could be introduced to avoid unwanted event detections in rapid succession. Alternatively, knowledge of previous events could be used to improve the prediction of future events, such as in an EMG-based spelling or a sign-language interface \cite{sharma_word_2014, trnka_user_2009}. 

\textbf{Wake words} \cite{kumar_verification-based_2021} are another method that could be used to greatly decrease false activations by enabling users to decide when they want to interact with the system or not. Similar to wake words for voice command interfaces, this wake word would be some form of highly recognizable EMG input (e.g., two quick finger snaps in succession), or even a voice or button-activated command. It is important to note, however, that wake words add an additional step to the system, reducing responsiveness and convenience, and do not explicitly solve the issue of misclassifications. Nevertheless, the use of wake words could alleviate the algorithmic challenge of distinguishing explicit inputs from general non-input muscle activity. This could be particularly beneficial in the early stages of developing robust EMG-based controls for general-purpose applications.

\subsection{User Training}
Although EMG-based interactions should arguably be intuitive and easy to learn, muscle-based control is still a novel input modality for most users. Generating separable and repeatable contractions resulting in EMG that a controller can easily classify can be challenging at first. As a result, some level of user training is crucial to ensuring that a user becomes capable of interacting through the given control system. Prosthetics researchers have investigated the effects of various training tools on learning EMG-based control \cite{dupont_myoelectric_1994,tabor_evaluation_2018,woodward_adapting_2019}. It has been shown that even amputees --- who are most often highly motivated to improve --- can find learning and adherence to training challenging \cite{winslow_mobile_2018}. This has led to a recent surge in the development of game-based and gamified EMG training tools to help motivate training \cite{tabor_designing_2017, winslow_mobile_2018, prahm_increasing_2017}. Unlike amputees, if general users cannot become competent with the given control system quickly they may not be sufficiently motivated to use it. Consequently, HCI researchers must understand the challenges around training, how it might impact overall control, and how to incentivize sufficient training within their applications. Whether through fun tutorials \cite{flatla_calibration_2011} or serious games \cite{tabor_designing_2017}, the training process should be an extension of the application. Furthermore, unlike traditional input modalities (e.g., keyboards), the cause of errors in pattern recognition-based myoelectric control can be difficult to understand since these systems are viewed as black boxes by most users (inputs go in and decisions come out). Therefore user training becomes critical to enable the possibility of experimentation in a controlled environment. Ultimately, EMG-based control is a motor skill, and as such, requires training to develop proficiency \cite{winslow_mobile_2018}.

\subsection{Adaptation} 
One of the limitations of EMG-based control systems is that their performance can degrade over time due to user fatigue, electrode shift, changes in user behaviours, and other transient factors. Static models trained on an initial training set that do not change over time to accommodate these changes are often at fault. This lack of robustness can lead to frustration and, inevitably, the abandonment of a device \cite{biddiss_upper_2007}. Adaptation can alleviate these issues by maintaining and improving the performance of systems by updating a model over time. Both supervised \cite{zhang_adaptation_2013, sensinger_adaptive_2009, cote-allard_deep_2019} and unsupervised \cite{cote-allard_unsupervised_2020,huang_novel_2017,sensinger_adaptive_2009} strategies have been explored for prosthesis control. Supervised adaptation requires additional oversight to label new ground truth training data, which is cumbersome for the user \cite{sensinger_adaptive_2009}. Because of this supervision, however, the classification accuracies for this method are often higher. Gamified calibration techniques, as proposed by \cite{flatla_calibration_2011}, could reduce the perceived effort. Unsupervised adaptation requires no user input, and data are tagged with pseudo-labels (i.e., predictions) through different automated strategies \cite{sensinger_adaptive_2009}. Due to its unsupervised nature, data mislabeling may occur and introduce errors into the model, possibly adding to the degradation over time. If properly constrained, such as through semi-supervision based on added application context (inferring what the user should have been doing), unsupervised adaptation would be the preferred approach for general-purpose applications. Interestingly, adaptation is not a new concept in HCI and has been explored for use cases such as adaptive interfaces \cite{mezhoudi_user_2013, eisenstein_adaptation_2000}, but it has not been widely leveraged in EMG works. 

\subsection{Activity Detection}
An arguably critical aspect of discrete control is activity detection; the extraction of a fragment of EMG corresponding to a dynamic gesture from a continuous stream of data. When classifying pre-segmented gestures in offline analyses, the extraction of relevant EMG activity is already done. However, during online use, data are continuously streamed to the system, and which segments of the signal stream to pass into a recognition model (i.e., where the start and end of a gesture are) must be determined. There are different techniques that have been proposed to solve this issue. When a period of inactivity is assumed to occur between gestures, \textbf{onset and offset detection} can be used to determine the onset/activation of EMG activity, denoting the beginning and ending of a gesture \cite{xu_adaptive_2013, pasinetti_novel_2015, merlo_fast_2003}. Another method employs \textbf{sliding windows}, in a process similar to the continuous windowing process, but at a larger time scale. In this scenario, larger windows (with a fixed length, determined based on assumed gesture lengths) are extracted, examined, and compared to target gestures. Although many such windows may not correspond to known gestures, thresholding techniques such as rejection may be used to avoid false activations. Nevertheless, more exploration is needed to determine the optimal strategies for extracting dynamic gestures associated with discrete events from a continuous stream of EMG.

\section{Discussion} \label{sec:section6}
This work has provided a new examination of EMG in HCI that recognizes the important influence of research and engineering in the field of prosthetics. Here we have characterized ``general-purpose applications'' as the use of EMG sensors as input to interactive systems, other than for the use of prosthetics. Our presentation thus far has demonstrated that while EMG work in prosthetics provides a good foundation, a deeper understanding, and a different focus are needed for EMG to be successful in HCI. Below, we critique our own work to better characterize its contribution.

Our exploration of the general-purpose use of EMG was intentionally focused. First, we focused primarily on HCI venues by focusing on the ACM DL rather than more broad searches or including other digital libraries. Second, while we followed a general procedure for collecting papers, the process was largely informed by the experience of our larger research group, which consists of two closely collaborating labs: one HCI lab in a computer science department; the other, a biomedical engineering lab collocated with a prosthetics clinic. Our groups have collaborated closely for several years, and so although our paper selection procedure was not a systematic treatment of the literature, we drew on the experiences of our larger research team that has straddled two distinct areas of research, making contributions to both HCI and prosthetics research. From this perspective, we believe we are uniquely positioned to characterize our observations and experiences. Nevertheless, our sampling of the literature was the source and motivation of this characterization, rather than simply being an organization of convenience. The exploration of the EMG literature in HCI, contrasted with that in the prosthetics field, enabled us to clearly characterize why HCI work and general-purpose applications have been falling short of their potential. It was through our organization and consideration of a larger body of work that we were able to identify common limiting factors. 

Our framework proposes a step-by-step process for the design and development of new EMG-based inputs for interactive systems. 
The framework is predominantly an expression and translation of the steps that prosthetics researchers employ in the development of a new EMG system, and, thus intuitive for us to identify. That being said, we have not found any source that has previously fully documented these steps in prosthetics or elsewhere. While this is a contribution in itself, we believe the novelty in this framework is in the succinct communication of the processes used in prosthetics for an HCI audience, allowing us to communicate how previous pitfalls can be avoided. Thus, we have provided a new and important starting point for any work in HCI that will employ EMG as an input modality.  

Interestingly, it was only after we had both fully described the themes from our exploration of previous work and enumerated the steps in our framework that we recognized the perfect alignment (the groupings of steps in our framework aligned with the themes identified from our exploration). This provides further confidence that the results of the exploration and the framework provide good coverage of the main concerns in developing EMG-based input. 

In this paper, we have necessarily focused on how the field of prosthetics has and can inform work in HCI, but we have largely ignored how HCI can inform and play a role in prosthetics research and practice. From improved prosthetics training \cite{tabor_designing_2017} to empowering amputees to create their own assistive tools \cite{hofmann-prototyping}, to a better understanding of how people value their prosthetic devices \cite{bennett-intimate}, the breadth of HCI research has already made an important impact and still has much to offer the field of prosthetics. We believe these two fields offer an ideal area for continued cross-fertilization that we have only begun to explore. In the next section, we make a call to researchers in both prosthetics and HCI to combine their expertise to unlock the potential of EMG as an important enabling technology that can be used by all as part of their everyday interactive systems. 

As emphasized throughout this work, EMG-based control research has primarily occurred within the prosthetics community, heavily influencing its development in HCI for general-purpose use. While this has been a good place to start, based on an understanding of the types of EMG control systems we need in this field, we now need to move in new directions. Current EMG control systems are designed for mechanically complicated multi-articulated prosthetic devices and for a user base with individual and potentially very different abilities from one another. In many general-purpose applications, the context is very different and the motivations, usage contexts, hardware, and technical challenges are unique. Instead of forcing our interactions into the prosthesis-control framework, we must explore other opportunities and avenues. For example, what algorithms and device form factors allow people to most effectively interact with mobile computing devices while exercising? Or, what gesture set provides the most expressive vocabulary for a virtual reality system? Answering these questions requires an evolution in how HCI researchers have been approaching EMG-based control. In fact, it is quite possible that the research field surrounding EMG-based control for general-purpose applications diverges from prosthetic applications. And, like in biomedical engineering, HCI researchers could dedicate substantial time to understanding and improving EMG control, including the development of conferences or publication venues focusing exclusively on the topic (e.g., \cite{mec_web}).


\subsection{A New Era of EMG Wearables and Beyond}
When the idea of EMG wearables was explored in the early 2000s \cite{saponas_demonstrating_2008}, sEMG technology was only beginning to become sufficiently inexpensive and accessible for general-purpose use. The 2014 release of the Myo Armband by Thalmic Labs revolutionized the space of EMG wearables \cite{rawat_evaluating_2016}, albeit not without its limitations. Firstly, its sampling rate was capped at 200 Hz, thus limiting its overall performance and robustness. Secondly, users did not always find it comfortable. Over long periods, people found it sweaty and heavy; and, for others, it could not be made to fit tightly enough for the electrodes to make consistent contact \cite{tabor_designing_2017}. Despite these limitations, the subsequent discontinuation of the Myo Armband has likely slowed the exploration of EMG for general-purpose applications because there is currently no relatively robust, low-cost device that is as readily available. However, ongoing commercial initiatives are trying to improve on past limitations (e.g., \cite{biomed, sifi_labs_web, mudra, jaloza_inside_2021}) to provide more robust, low-cost wearable EMG devices.

EMG lends itself to wearable form factors because of its small size and low power requirements. For an EMG wearable to be successful it can be assumed that it should be accessible, comfortable, convenient, and easy to use. Following this, wearables should, therefore, be available in a range of forms (e.g., armbands, watches, or bracelets) to enable the most convenient and effective format for a particular context. They should also have sufficient electrodes (i.e., 6-10) with a sampling rate of at least 500 Hz, to allow for the robust detection of complicated inputs \cite{hargrove_comparison_2007,li_selection_2010}. With that said, each electrode increases the required circumference of a wearable device. As such, armbands should be adjustable or come in different sizes. It is also important that these wearable devices have built-in IMUs (Inertial Measurement Units) to allow for improved control through movements combined with EMG (e.g., for controlling a cursor \cite{haque_myopoint_2015}), but also to improve the accuracy of EMG-based input (e.g., arm position affects the production of EMG signals \cite{campbell_current_2020,radmand_suitability_2014}). Exploring this combination of EMG and IMU as complimentary sensing modalities is an exciting direction of potential future research. Additionally, the continued research and development of pre-built user-independent and context-independent models (for both static contractions and dynamic gestures) could further facilitate the integration of EMG in new applications. 

Beyond what we traditionally think of as EMG wearables (e.g., cuffs or wristbands), an increasing body of prosthetics research is dedicated to other EMG measurement modalities (as opposed to sEMG). Invasive techniques, such as implanted and intramuscular electrodes, can vastly improve the quality of EMG signals \cite{hahne_novel_2016} and thus improve the usability of such systems \cite{kamavuako_usability_2014, smith_comparison_2013, cipriani_dexterous_2014}. Alongside other surgical augmentation procedures such as osseointegration \cite{albrektsson_osteoinduction_2001, mastinu_embedded_2017} and targeted muscle reinnervation \cite{kuiken_targeted_2009}, these techniques can vastly improve the quality and robustness of prosthesis control. Given the research and public interest in body augmentation \cite{britton-cyborg}, body implants \cite{strohmeier-implants}, and other neurological links \cite{Frauenberger-entanglement}, future work could explore the reception and improved performance of interfaces other than generic sEMG in general-purpose applications.

Similarly, EMG inputs could be combined with EMS (electrical muscle stimulation; sometimes referred to as functional electrical stimulation, or FES, in biomedical engineering \cite{nithin}). EMS has been widely explored in recent HCI work as a form of output (e.g., \cite{ems-io, ems-toolkit}, and is effectively the opposite of EMG recordings (instead of measuring the electrical activity of contractions, EMS stimulates muscle contractions using electrical activity). The notion of having a wearable device that enables both input and output through a muscle interface is compelling and has been successfully demonstrated by Duente et al. \cite{ems-io}. Future work could explore how such technologies could enable richer forms of input/output built into a single wearable device. 


\subsection{Facilitating EMG-Based Development and Prototyping}
One of the major barriers limiting the development of EMG-based control is its inherent complexity. Ideally, a toolkit would exist that facilitates the design and development of EMG-based interactions using state-of-the-art control systems. Such a toolkit would have the goal of adding EMG-based control to an interactive system with as close to a plug-and-play experience as possible. This would enable HCI researchers to focus on the interaction rather than on the design of the control system. For the success of the toolkit, however, it would be imperative that it was validated by domain experts (e.g., EMG researchers from both HCI and prosthetics). Additionally, it should be well-documented with examples and data sets  --- including both static contractions and dynamic gestures --- for facilitating the iterative design process. One challenge for such a toolkit is that EMG-based technology and control systems are continuously evolving. As such, this toolkit should be open-sourced and ideally adopted by a community of researchers to allow it to evolve. EMBody \cite{karolus_embody_2021} is an excellent initial effort towards such a toolkit. However, EMBody is limited by many of the issues explored in this paper. For example, it does not support the use of discrete control systems, thus unintentionally influencing researchers to follow a continuous control framework when designing their system. Ultimately, this is one of the important barriers to the adoption we discussed in the emerging themes section (see section \ref{sec:InteractionDesign}). From a data perspective, the Nina Pro Database \cite{atzori_building_2012} highlights the importance of having a robust open-sourced dataset. A fully featured toolkit and corresponding dataset would not only enable more researchers to experiment with the technology but it could be crucial to the future success of EMG.

\subsection{Ease of Use and Convenience}
EMG has become the dominant interface for prosthesis use as it enables device control by solely relying on input from the residual muscles --- those remaining above the amputation. In most cases, amputees do not have access to other control approaches --- such as computer vision, typical controllers, or data gloves --- to control their prostheses. Since EMG is currently the main viable option, it has inevitably been adopted by amputees. However, users interacting with general-purpose applications will likely have the option of using these other technologies. As such, EMG should provide benefits that others cannot, including increased convenience and ease of use. Individuals should want to use EMG as an input modality, not because it is novel but because it is discrete, robust, reliable, and easy to use. One crucial step in making this a reality is by focusing on robust user- and context-independent models. Users will not be keen if they have to tediously re-train a model multiple times per day or every time the specific context changes. In theory, these systems will be plug-and-play with the option of allowing additional calibration and customization if desired for more efficient interactions, such as with modern speech recognition engines. Finally, these systems should be as intuitive as possible to reduce the training time required to achieve control proficiency. Addressing these concerns will drastically improve usability and hopefully in turn, the adoption of this technology.

\subsection{Discrete and Continuous Input Discovery}
One of the major contributions of this work was differentiating discrete (i.e., dynamic gestures) and continuous (i.e., static contractions) inputs for EMG-based interactions. In doing so, we have highlighted a crucial need for evaluating these two groups of inputs separately. Static contractions have dominated the research space due to their prevalence in prosthesis control. As such, physiologically appropriate contractions --- those consistent with the prosthetic components they are meant to control --- have been the primary focus. For similar reasons, dynamic gestures have been largely neglected in EMG-based prosthesis control. From an HCI perspective, however, this motivation may be reduced, and other contractions and gestures may be preferred. As such, work must explore what gestures can be recognized and the best methods for recognition. In this exploration, the IMU becomes appealing as it adds additional information for classifying dynamic movements --- understanding these impacts is crucial. For example, one research goal could be to understand what gestures can be recognized robustly with and without the IMU when combined with EMG. Or potentially, researchers could explore how EMG could be leveraged in addition to traditional computer vision or data glove approaches. We must also explore gesture sets that are less likely to cause false activations. Gestures should be unique enough from common muscular contractions to reduce false activations as users participate in other activities \cite{chang_wearable_2020}. Another interesting consideration arises in scenarios where subtle gestures are desirable and that are not conducive to large dynamic gestures (e.g., dismissing a call during a meeting). Is there a set of gestures that are distinguishable, yet discreet enough for these contexts? Future work should explore the identification of standardized inputs (including both static contractions and dynamic gestures) that are intuitive, robust, and ideal for general-purpose use.

\section{Conclusion}
With this paper, we aim to deepen the understanding and research practices around a frequently explored input technology, EMG-based sensors. As computing continues to move away from desktop settings, EMG offers potentially low-cost, small, and wearable sensors that can fit into a wide range of scenarios where other technologies often do not work well. Despite its long history of use in prosthetics, we are still without an application or consumer-grade technology that has revolutionized the use and uptake of EMG for general use; despite the notable past and ongoing efforts. We believe that it is not because it is impossible, but instead that more coordinated research and development is needed to realize truly robust, reliable, and intuitive EMG input. Towards this end, we demonstrate fundamental limitations to previous applications of EMG-based input, and, through our design framework, we provide a starting point for HCI researchers to better design successful EMG-based interactions. The framework draws on the long history of EMG-based prosthetics research and distinguishes the main concepts that can be applied to general-use applications. Finally, we call the HCI community to action by providing a research agenda for HCI researchers interested in EMG-based control. The new directions for research that we identify are somewhat distinct from those in prosthetics. While HCI and the prosthetics field can learn from one another, there will inevitably be a need for some divergence as the HCI community's understanding and practices around EMG continue to mature, and we unlock the hidden potential of EMG as an input technology for a wide range of interactive systems.

\begin{acks}
We want to thank all of the experienced staff, researchers and clinicians, past and present, at the Institute of Biomedical Engineering and the Human Computer Interaction Lab at the University of New Brunswick who provided their guidance and expertise in making this project possible. This research was supported in part by the Natural Sciences and Engineering Research Council of Canada (NSERC) and the Atlantic Canada Opportunities Agency.
\end{acks}

\bibliographystyle{ACM-Reference-Format} 
\bibliography{bibliography}

\nociteR{ke_intersected_2020_1}
\nociteR{kim_emg-based_2008_1}
\nociteR{javaid_comparative_2018_1}
\nociteR{zhang_hand_2009_1}
\nociteR{eghtebas_investigation_2018_1}
\nociteR{naik_hand_2006_1}
\nociteR{benatti_towards_2014_1}
\nociteR{vasiljevas_development_2014_1}
\nociteR{delpreto_plug-and-play_2020_1}
\nociteR{huang_leveraging_2015_1}
\nociteR{karolus_emguitar_2018_1}
\nociteR{amma_advancing_2015_1}
\nociteR{saponas_enabling_2009_1}
\nociteR{saponas_making_2010_1}
\nociteR{saponas_demonstrating_2008_1}
\nociteR{costanza_toward_2005_1}
\nociteR{costanza_intimate_2007_1}
\nociteR{hu_comprehensive_2020_1}
\nociteR{dai_capg-myo_2021_1}
\nociteR{zadeh_evaluating_2018_1}
\nociteR{tsuboi_proposal_2017_1}
\nociteR{mcintosh_empress_2016_1}
\nociteR{kerber_same-side_2015_1}
\nociteR{robinson_pattern_2017_1}
\nociteR{koskimaki_myogym_2017_1}
\nociteR{sueaseenak_optimal_2017_1}
\nociteR{mulling_characteristics_2015_1}
\nociteR{haque_myopoint_2015_1}
\nociteR{karolus_hit_2020_1}
\nociteR{robinson_effectiveness_2018_1}
\nociteR{assad_biosleeve_2013_1}
\nociteR{asai_finger_2019_1}
\nociteR{becker_touchsense_2018_1}
\nociteR{tortora_dual-myo_2019_1}
\nociteR{touyama_prototype_2005_1}
\nociteR{al-jumaily_electromyogram_2009_1}
\nociteR{ho_myobuddy_2017_1}
\nociteR{sharma_neural_2017_1}
\nociteR{caramiaux_understanding_2015_1}
\nociteR{paudyal_sceptre_2016_1}
\nociteR{li_automatic_2010_1}
\nociteR{zhang_intelligent_2022_1}
\nociteR{zhang_application_2022_1}
\nociteR{kerber_user-independent_2017_1}
\nociteR{choi_preliminary_2019_1}
\nociteR{karolus_embody_2021_1}
\nociteR{stoica_remote_2014_1}

\bibliographystyleR{ACM-Reference-Format}
\bibliographyR{reviewed_papers}
\label{ReviewedWork}

\end{document}